\documentclass[reprint,amsmath,amssymb,aps,]{revtex4-2}

\usepackage{graphicx, commath,amsfonts}
\usepackage{xcolor, ulem}
\usepackage{dcolumn}
\usepackage{bm}
\newcommand{\iu}{\mathrm{i}\mkern1mu}
\begin{document}
	
	
	\title{Sakaguchi Swarmalators}
	
	\author{Joao U.F. Lizárraga}
	\author{Marcus A.M. de Aguiar}%
	\email{aguiar@ifi.unicamp.br}
	\affiliation{%
		Instituto de Física Gleb Wataghin, Universidade Estadual de Campinas, Unicamp 13083-970, Campinas, São Paulo, Brazil
	}
	
	\date{\today}
	
\begin{abstract}
		Swarmalators are phase oscillators that cluster in space, like fireflies  flashing on a swarm to attract mates. Interactions between particles, which tend to synchronize their phases and align their motion, decrease with the distance and phase difference between them, coupling the spatial and phase dynamics. In this work, we explore the effects of disorder induced by phase frustration on a system of Swarmalators that move on a one-dimensional ring. Our model is inspired by the well-known Kuramoto-Sakaguchi equations. We find, numerically and analytically, the ordered and disordered states that emerge in the system. The active states, not present in the model without disorder, resemble states found previously in numerical studies for the 2D Swarmalators system. One of these states, in particular, shows similarities to turbulence generated in a flattened media. We show that all ordered states can be generated for any values of the coupling constants by tuning the phase frustration parameters  only. Moreover, many of these combinations display  multi-stability.
\end{abstract}

\maketitle

\section{\label{sec:level1}Introduction}

Synchronization and swarming are emergent phenomena observed in various living systems. The former refers to the tendency of individuals' states to converge towards specific periodic behaviors and has been widely investigated using the  Kuramoto~\cite{acebron2005kuramoto,Rodrigues2016} or the Stuart-Landau \cite{matthews1991dynamics,aoyagi1995network} models. The latter describes systems in which individuals tend to aggregate and align in space, as often observed in animals such as birds and fish~\cite{katz2011inferring, cavagna2023natural}. Although the two behaviors have been spotted independently in nature, systems including the Japanese tree frogs~\cite{aihara2014spatio} and the Quincke rollers~\cite{zhang2020reconfigurable}, among others~\cite{giomi2013swarming, tan2022odd, creppy2016symmetry}, suggest that synchronization and swarming also occur together. A model that couples both behaviors was recently proposed in \cite{o2017oscillators} and the corresponding particles termed swarmalators. 

The Swarmalators model~\cite{o2017oscillators} describes a system of particles characterized by internal phases $\theta_i$ and spatial positions $\vec{x}_i$. Phase and position dynamics coupled in such a way that phases tend to synchronize among nearby particles and velocities tend to align more easily among particles with synchronized phases. An instance of the model, for $N$ particles moving in a two-dimensional space, is described by
\begin{align}\label{eq:2D}
	\begin{split}
		\dot{\vec{x}}_{i}&= \frac{1}{N} \sum_{j\neq i}^{N} \bigg[\frac{\vec{x}_{j} - \vec{x}_{i}}{|{\vec{x}_{j} - \vec{x}_{i}}|} (1+ J\cos(\theta_{j}- \theta_{i}))- \frac{\vec{x}_{j} - \vec{x}_{i}}{|{\vec{x}_{j} - \vec{x}_{i}}|^2} \bigg] \\ 
		\dot{\theta}_{i}&=  \frac{K}{N} \sum_{j\neq i}^{N} \frac{\sin(\theta_{j}- \theta_{i})}{|\vec{x}_{j}- \vec{x}_{i}|}.
	\end{split}
\end{align}

It has been shown that different collective states may emerge for specific sets of parameters $K$ and $J$ \cite{o2017oscillators}. Previous work have also explored the system's behavior under external stimulus~\cite{lizarraga2020synchronization}, variations on the nature of individual's interactions~\cite{hong2018active, sar2022swarmalators, lee2021collective, jimenez2020oscillatory, japon2022intercellular} and effects of thermal noise \cite{hong2023swarmalators}. However, from an analytical perspective almost no progress has been made. Under this premise, O'Keeffe et. al~\cite{o2022collective} proposed a one-dimensional analogue of the model whose dynamics are governed by
\begin{align}
	\begin{split}
		\dot{x_{i}} = \frac{J}{N} \sum_{j}\sin(x_{j} - x_{i})\cos(\theta_{j} - \theta_{i}),\\
		\dot{\theta_{i}} = \frac{K}{N} \sum_{j}\sin(\theta_{j} - \theta_{i})\cos(x_{j} - x_{i}),
	\end{split}
	\label{eq:ring}
\end{align}
and capture several features of Eqs.~\eqref{eq:2D}. This simpler model displays the emergence of several static collective states observed in the 2D system and, most importantly, can be treated analytically. Further work have also considered noisy interactions~\cite{hong2023swarmalators}, distributed couplings~\cite{o2022swarmalators}, random pinning~\cite{sar2023pinning}, and intrinsic oscillating frequencies~\cite{yoon2022sync}.

The 1D Swarmalators model, however, cannot describe some of the active states displayed by the full 2D system. Even if some states of the 2D model are arranged in the form of an annulus, projecting it onto a 1D ring leaves out part of the dynamics that could be essential for the formation of the structure. On the other hand, the similarity of the 1D model with a pair of coupled Kuramoto equations, suggests that the expertise acquired from studying this famous synchronization model can be leveraged to analyze Swarmalators systems of this type. Here we propose a model of frustrated 1D swarmalators, based on the Sakaguchi-Kuramoto model~\cite{sakaguchi1986soluble}, as a source of disorder that could compensate for the loss of freedom of the 1D system and potentially restore the active states observed in the 2D model. We call the corresponding particles {\it Sakaguchi Swarmalators}.   We explore the effects of additional phase frustration parameters to both the spatial and phase dynamics in Eqs.~\eqref{eq:ring}. This type of disorder differs from (and complements) that produced by distributed couplings, studied by several authors for the Kuramoto model~\cite{acebron2005kuramoto}, and by O'Keeffe and Hong for the 1D Swarmalators model~\cite{o2022swarmalators}. We will show that, indeed, frustration leads to coherent active states in 1D, similar to the ones found in the 2D setup. Additionally, it is worth mentioning that the new parameters of the Sakaguchi Swarmalators model have similarity with ``offset terms'' used in a recent numerical study of a modified 2D Swarmalators system~\cite{ceron2023diverse}. 

We describe the modifications to the original 1D model in (Sec.~\ref{sec:model}), introducing disorder as in the Kuramoto-Sakaguchi model. Then, in the same section, we present the different collective states obtained from numerical computations. In Sec.~\ref{sec:stab_analysis}, we present the stability analyses of states that show ordered configurations. The conditions obtained from the analytical computations allow us to picture the stability regions in the space of frustration parameters (Sec.~\ref{sec:stab_maps}). Finally, in Sec.~\ref{sec:concl} we sum up some concluding remarks.

\section{The Sakaguchi Swarmalators model}
\label{sec:model}
Our modifications to the 1D Swarmalators model bring back the essential feature of the Kuramoto-Sakaguchi model: the introduction of disorder on a system intended to exhibit a coherent behavior~\cite{sakaguchi1986soluble}. In our model, however, the disorder affects both the spatial and phase components of each particle's dynamics. As described by the expressions
\begin{align}
	\begin{split}
		\dot{x_{i}} = \frac{J}{N} \sum_{j}\sin(x_{j} - x_{i} + \alpha)\cos(\theta_{j} - \theta_{i} + \beta),\\
		\dot{\theta_{i}} = \frac{K}{N} \sum_{j}\sin(\theta_{j} - \theta_{i} + \beta)\cos(x_{j} - x_{i} + \alpha)
	\end{split}
	\label{eq:frust_model}
\end{align}
so that the disorder parameters $\alpha$ and $\beta$ are incorporated to the system dynamics. Hence, if the system reaches coherence in phase ($\theta_j \approx \theta_i$) or space ($x_j \approx x_i$), the effects induced by $\alpha$ and $\beta$ would lead to its disruption.

\subsection{Equilibrium states}

\begin{figure*}
	\centering
	\includegraphics[width = 0.9\textwidth]{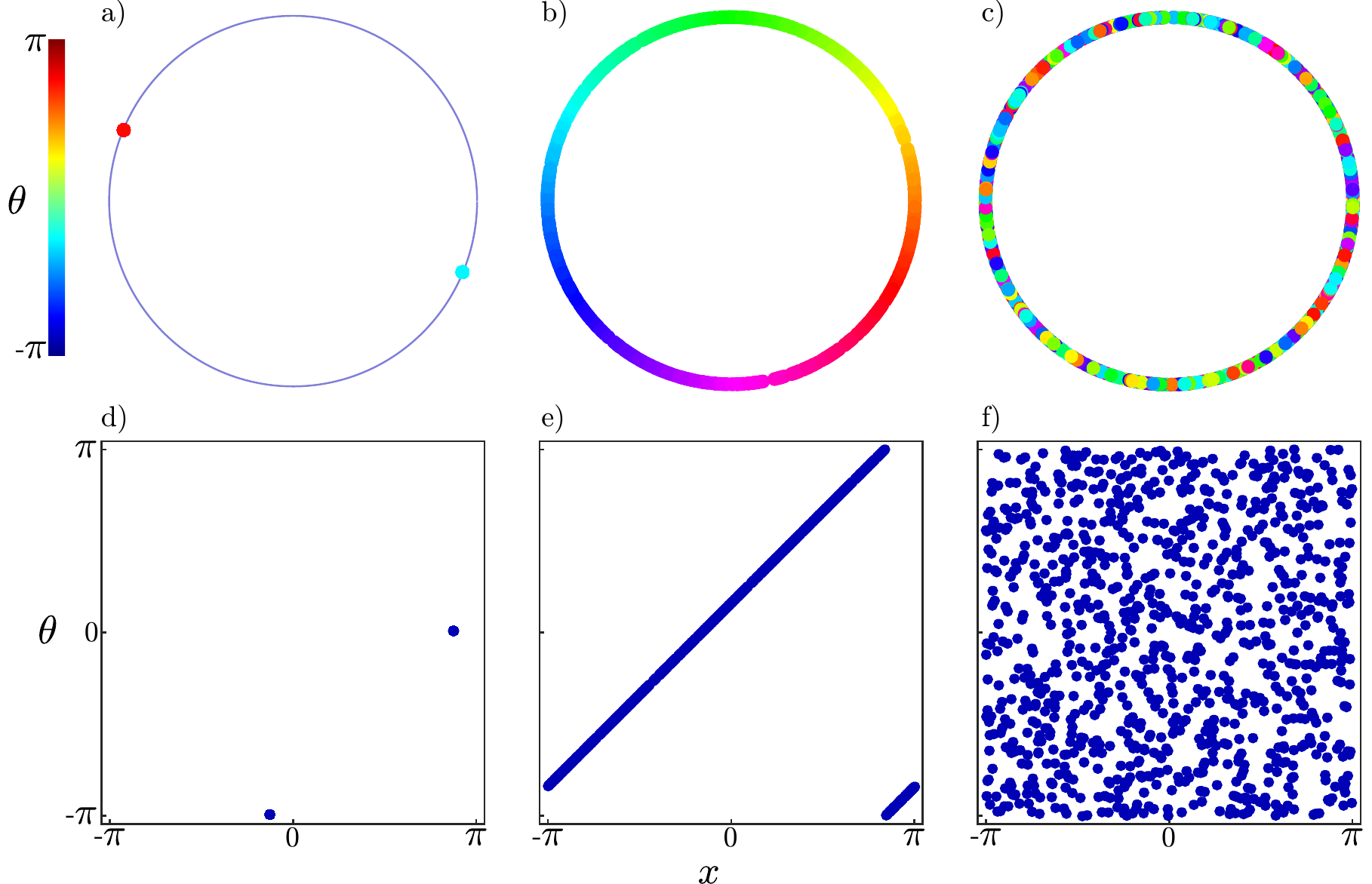}
	\caption{Spatial behavior (top row) and position-phase correlations (bottom row) for the static states of the Sakaguchi Swarmalators model. ($K = 1$) is set for all the simulations, and ($\gamma^+$, $\gamma^-$) are set as a,d) ($0$, $0$) for the Static Synchronous, b,e) ($1.67$, $0$) for the Static Phase Wave, and c,f) ($1.8$, $2$) for the Static Asynchronous states.}
	\label{fig:static_states}
\end{figure*}

\begin{figure*}
	\centering
	\includegraphics[width = 0.9\textwidth]{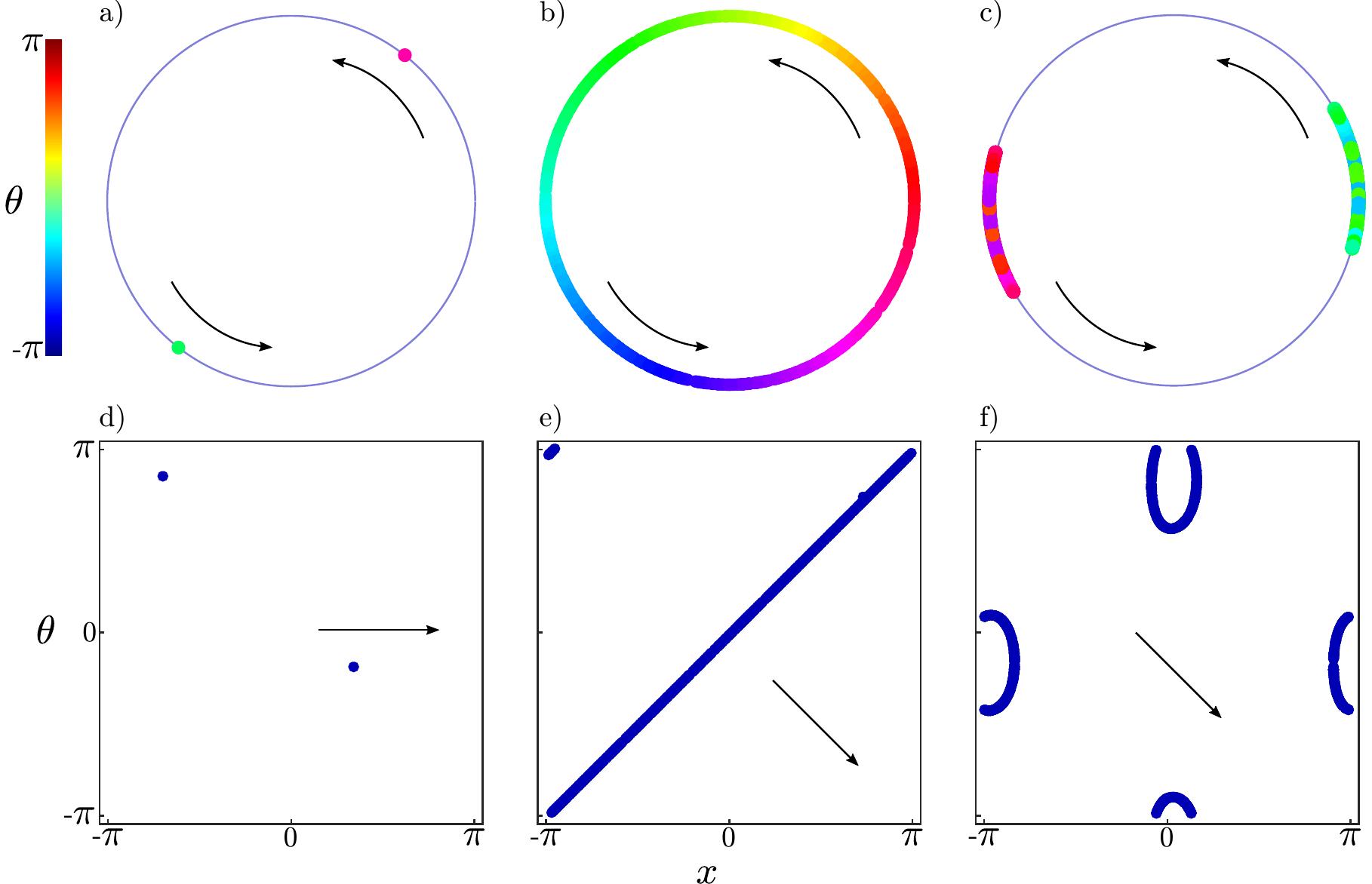}
	\caption{Spatial behavior (top row) and position-phase correlations (bottom row) for the ordered active states of the Sakaguchi Swarmalators model. ($K$, $\gamma^+$, $\gamma^-$) are set as a,d) ($1$, $1.3$, $1.5$) for the Active Synchronous, b,e) ($1$, $2.5$, $1.1$) for the Active Phase Wave, and c,f) ($-5$, $0.3$, $-3.1$) for the Ring states (see movies S1, S2, and S3 in Supplemental Material). Arrows represent the translation direction of the particles.}
	\label{fig:o_active_states}
\end{figure*}

\begin{figure*}
	\centering
	\includegraphics[width = 0.9\textwidth]{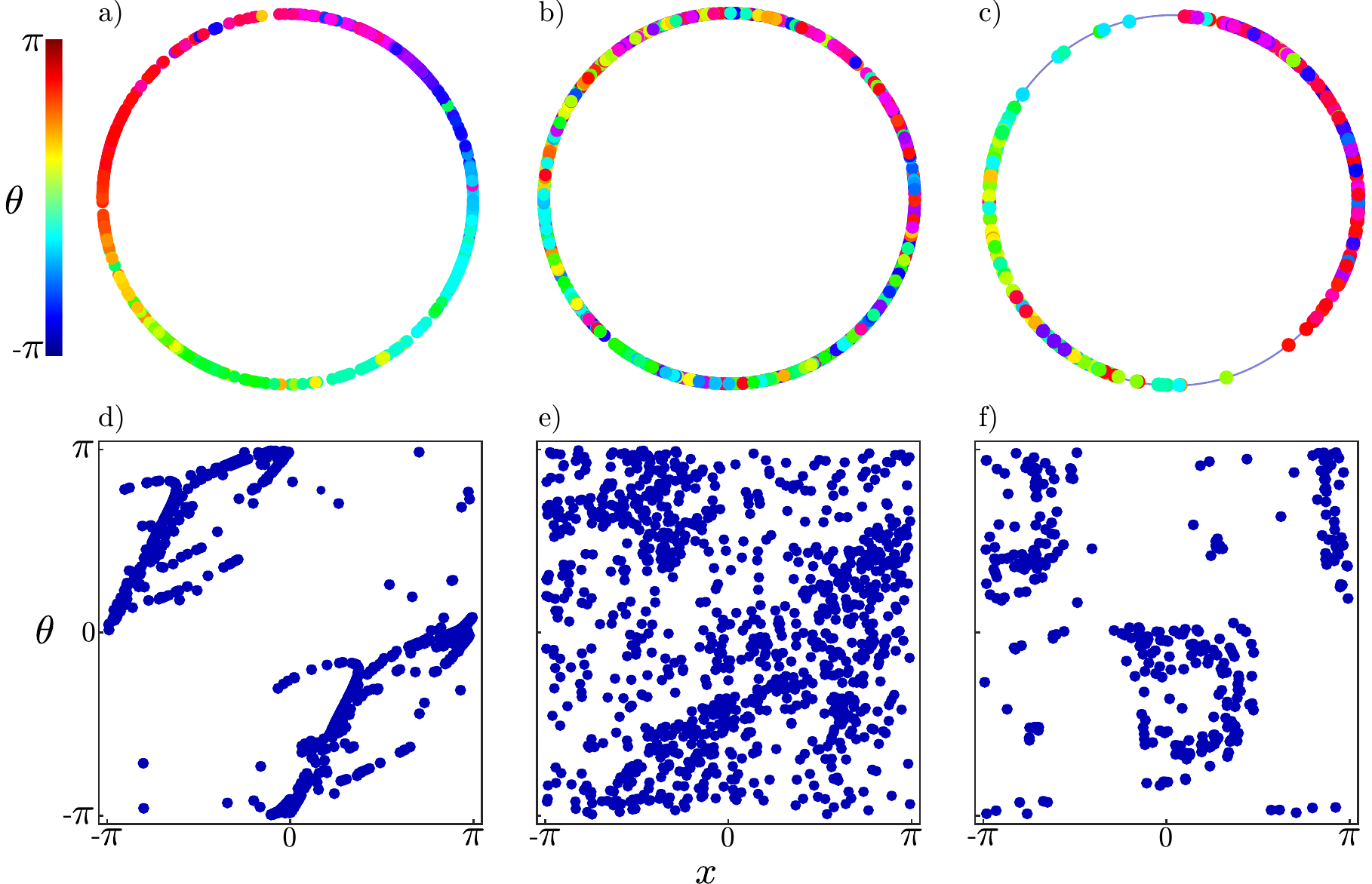}
	\caption{Spatial behavior (top row) and position-phase correlations (bottom row) for the disordered active states of the Sakaguchi Swarmalators model. ($K$, $\gamma^+$, $\gamma^-$) are set as a,d) ($-0.2$, $1.9$, $-1.4$) for the Noisy Active Phase Wave, b,e) ($-1$, $1.25$, $-0.25$) for the Active Asynchronous, and c,f) ($-0.2$, $3$, $-0.5$) for the Turbulent states (see movies S4, S5, and S6 in Supplemental Material). In all cases, particles are in continuous motion.}
	\label{fig:d_active_states}
\end{figure*}

Numerical computation of Eqs.~\eqref{eq:frust_model} allows us to get insights on its long term behavior. Snapshots of the system's collective behavior, after $10^4$ time steps, are shown in Figures~\ref{fig:static_states},~\ref{fig:o_active_states}, and~\ref{fig:d_active_states}. In all cases, the number of particles is $N = 500$ and the parameter $J$ is set to $1$. Thus, the control parameters are $K$, $\alpha$, and $\beta$. In the figures we use the auxiliary parameters $\gamma^\pm = \alpha \pm \beta$, instead, since $\gamma^\pm$ will be relevant for the stability analyses in the next sections. Particles' positions and phases are initially distributed uniformly in ranges $-\pi$ to $\pi$.  In the spatial pictures, shown in the top rows of Figures~\ref{fig:static_states},~\ref{fig:o_active_states}, and~\ref{fig:d_active_states}, particles are positioned along the ring and colored according to their phases. The scatter plots, in the bottom row of Figures~\ref{fig:static_states},~\ref{fig:o_active_states}, and~\ref{fig:d_active_states}, correspond to the spatial picture above and show the position-phase correlation for each particle. The states presented in Figure~\ref{fig:static_states} demonstrate the convergence of particles to fixed values in phase and space. Once the particles reach these states, after a transient period, they remain there statically. Despite this feature, however, we observe clear differences in the position-phase coherence of each case. The Static Synchronous state (Figures~\ref{fig:static_states}a and~\ref{fig:static_states}d) shows the formation of two clusters spaced, in phase and space, by a factor of $\pi$. Particles move to each cluster depending on their initial condition and synchronize with its cluster neighbors. In the Static Phase Wave state, (Figures~\ref{fig:static_states}b and~\ref{fig:static_states}e), particles are uniformly distributed along the ring. Moreover, each particle's phase is correlated with its position, implying that these are also distributed uniformly. The correlation shown in the figure is positive, however, variations in the initial conditions can change the behavior of the system so that the steady behavior leads to a negative correlation. The Static Asynchronous state (Figures~\ref{fig:static_states}c and~\ref{fig:static_states}d) shows that particles are distributed uniformly in phase and space. However, unlike the Static Phase Wave state, particles' positions and phases are uncorrelated.

Active analogs of the Static Synchronous and Static Phase Wave states are shown in Figure~\ref{fig:o_active_states}. In the Active Synchronous state (Figures~\ref{fig:o_active_states}a and~\ref{fig:o_active_states}d) the two clusters of particles, as described before for the Static Synchronous case, are rotating along the ring. Despite the rotation, the clusters preserve the spacing of $\pi$ in position and phase. A similar effect is seen in the Active Phase Wave state (Figures~\ref{fig:o_active_states}b and~\ref{fig:o_active_states}e), where the uniformily distributed particles rotate while keeping the position-phase correlation. In Figures~\ref{fig:o_active_states}c and~\ref{fig:o_active_states}f, we introduce a new state where particles rotate around the ring while keeping a fixed pattern. In this state, particles cluster on a position-phase region, in contrast to the Active Synchronous state, where particles cluster on two $\pi$-distanced points.

In Figure~\ref{fig:d_active_states}, we show three additional active states. Despite not being completely ordered these states still show the emergence of intriguing patterns. In the Noisy Active Phase Wave state (Figures~\ref{fig:d_active_states}a and~\ref{fig:d_active_states}d), particles move and initially form a correlated position-phase pattern. After some time, however, this coherence is destroyed and a dynamic behavior starts where distorted correlation appears and disappears continuously. The Active Asynchronous state (Figures~\ref{fig:d_active_states}b and~\ref{fig:d_active_states}e) is the active analogous to the Static Asynchronous state, shown in Figures~\ref{fig:o_active_states}c and~~\ref{fig:o_active_states}e. In this state, however, particles jiggle and move randomly. The effects of this dynamic behavior, as better shown in the state's scatter plot, generate position-phase correlation in a non-uniformly distributed configuration. The last disordered state (Figures~\ref{fig:d_active_states}c and~\ref{fig:d_active_states}f) is named Turbulent and is unrelated to the previous ones. In this state, the particles move randomly along the ring without an specific position-phase coherence. However, as presented on its scatter plot, a recurrent pattern emerges where the particles' position-phase correlations generate vortexes that rotate and move around while exchanging individuals. 

The figures described in this section, obtained numerically, allow us to understand the coherent and incoherent behaviors of the Sakaguchi Swarmalators system. In the next section we explore the analytical features of the model, describing some of these states and their stability conditions. 
\section{Stability analyses}
\label{sec:stab_analysis}

Following \cite{o2022collective} we define $\xi_{i} = x_{i} + \theta_{i}$, $\eta_{i} = x_{i} - \theta_{i}$ and rewrite Eqs.~\eqref{eq:frust_model} as
\begin{align}
	\begin{split}
		\dot{\xi}_{i} &= \frac{J_{+}}{N}\sum_{j}\sin(\xi_{j} - \xi_{i} + \gamma^+) + \frac{J_{-}}{N}\sum_{j}\sin(\eta_{j} - \eta_{i} + \gamma^-),\\
		\dot{\eta}_{i} &= \frac{J_{-}}{N}\sum_{j}\sin(\xi_{j} - \xi_{i} + \gamma^+) + \frac{J_{+}}{N}\sum_{j}\sin(\eta_{j} - \eta_{i} + \gamma^-),    
	\end{split}
	\label{eq:rew_frust_model}
\end{align}
where $J_\pm =(J\pm K)/2$ and  $\gamma^\pm = \alpha \pm \beta$. We also define the order parameters 
\begin{align}
	\begin{split}
		S_+ e^{\iu\phi_+} &=\frac{1}{N}\sum_{j}e^{\iu \xi_{j}},\\
		S_- e^{\iu\phi_-} &=\frac{1}{N}\sum_{j}e^{\iu \eta_{j}},
	\end{split}
	\label{eq:OP_fin}
\end{align}
where the real values $S_\pm$ (ranging from $0$ to $1$) are coherence metrics associated with positive or negative correlations between particles' positions and phases. For instance, the coherence of the Static Phase Wave state, shown in Figure~\ref{fig:static_states}b, is $S_+ \approx 1$ due to the position-phase positive correlation of the particles. For the Static Asynchronous state, shown in Figure~\ref{fig:static_states}c, on the other hand, $S_\pm \approx 0$, since there is no correlation between particles' positions and phases.

\subsection{Synchronous states}\label{sub:SS}
These states involve clustering and synchronization of the particles, that converge simultaneously to specific values in phase and space that can be static or dynamic. Setting $\xi_j=\xi_i =\xi$ and $\eta_j = \eta_i = \eta$ in Eqs.~(\ref{eq:rew_frust_model}) we obtain the equilibrium trajectories 
\begin{align*}
	\xi & = J_{+}\sin(\gamma^+)t + J_{-}\sin(\gamma^-)t + \xi_0,\\
	\eta & = J_{-}\sin(\gamma^+)t + J_{+}\sin(\gamma^-)t + \eta_0.
\end{align*}
To analyze the stability of this solution, we add small perturbations $\delta\xi_i$ and $\delta\eta_i$ to each particle around the equilibrium trajectory and compute their dynamic behavior. The temporal evolution of the perturbations is described by  
\begin{widetext}
	\begin{align}
		\begin{split}
			\delta\dot{\xi_{i}} &= J^+\cos(\gamma^+)\sum_{j}(\delta\xi_{j} - \delta\xi_{i}) +J^-\cos(\gamma^-)\sum_{j}(\delta\eta_{j} - \delta\eta_{i}) \\
			\delta\dot{\eta_{i}} &= J^-\cos(\gamma^+)\sum_{j}(\delta\xi_{j} - \delta\xi_{i}) +J^+\cos(\gamma^-)\sum_{j}(\delta\eta_{j} - \delta\eta_{i}), 
		\end{split}
		\label{eq:ss_perturb}
	\end{align}
\end{widetext}

where $J^\pm = J_{\pm}/N$. These equations form a $2N\times2N$ linear system which is evaluated in detail in Appendix~\ref{sec:appendix}. The eigenvalues, that determine the stability of the equilibrium trajectory, are
\begin{widetext}
	\begin{align}
		\begin{split}
			\lambda_\pm^{SS} &= -\frac{J_+}{2}\left(\cos({\gamma^+}) + \cos({\gamma^-})\right)\pm\frac{1}{2}\left({J_+}^2\left(\cos({\gamma^+}) - \cos({\gamma^-})\right)^2 + 4{J_-}^2\cos({\gamma^+}) \cos({\gamma^-})\right)^{1/2},\\
			\lambda_0^{SS} &= 0,
		\end{split}
		\label{eq:eigen_SS}
	\end{align}
\end{widetext}
where $\lambda_{\pm}^{SS}$ have multiplicity $(N-1)$ each, and $\lambda_0^{SS}$ has multiplicity $2$. The superscript $SS$ stands for Synchronous States.

For the particular case where $J = K$, and therefore $J_- = 0$, the non-zero eigenvalues are
\begin{align}
	\begin{split}
		\lambda_{+}^{SS(J = K)} &= -J_{+}\cos(\gamma^-),\\
		\lambda_{-}^{SS(J = K)} &= -J_{+}\cos(\gamma^+),
	\end{split}
	\label{eq:simplifySS}
\end{align}
and will be negative for $J_+ > 0$ when $\gamma^\pm\in[-\pi/2, \pi/2]$, and for $J_+ < 0$ when $\gamma^\pm\in[\pi/2, 3\pi/2]$. The stability regions for $J\neq K$ are more complicated due to the shape of the non-zero eigenvalues surfaces. We shown the stability regions in this case in the next section. 

Before we close this subsection we note an interesting symmetry that appears for $J = 1$. In this specific case, $J_\pm = (K \pm 1)/2$, and considering $K^* = 1/K$, we get $J_{\pm}^* = \pm J_{\pm}/K$. Then, if we find ${J_{\pm}^*}^2$ an plug it into Eqs.~\eqref{eq:eigen_SS}, we see that the non-zero eigenvalues will just be scaled as
\begin{equation}
	{\lambda_{\pm}^{SS}}^* = \frac{1}{K}\lambda_{\pm}^{SS},
	\label{eq:eig_scaling}
\end{equation}
so the stability regions for $K$ and $1/K$ are exactly the same.

	
	\subsection{Phase Wave states}\label{sub:PW} 
	Here the particles are distributed uniformly in space and phase but these variables are correlated. Also, they can move rigidly, keeping their relative positions and phases constant. These states are represented by $x_{i} = 2i\pi/N + x_{0} + v_x t$ and $\theta_{i} = \pm 2i\pi/N + \theta_{0} + v_\theta t$, where $v_x$ and $v_\theta$ can be determined from Eqs. (\ref{eq:rew_frust_model}). The $\pm$ sign in $\theta_{i}$ depends on the type of position-phase correlation. We consider a negative correlation, so that the equilibrium trajectories must satisfy
	\begin{align*}
		\xi_{i} & = J_{+}\sin({\gamma^+})t + \xi_{0},\\
		\eta_i &\ = J_{-}\sin({\gamma^+})t + \frac{4\pi i}{N} + \eta_{0}. 
	\end{align*}
	Substituting in Eqs.~(\ref{eq:rew_frust_model}) we find $v_x+v_\theta = J_+ \sin\gamma^+ $ and $v_x-v_\theta=J_ - \sin\gamma^+ $. To study the stability of this solution, we again add perturbations $\delta \xi_i$ and $\delta \eta_i$ to the equilibrium and find their dynamics. We obtain
	\begin{widetext}
		\begin{align}
			\begin{split}
				\delta\dot{\xi_{i}} &= J^+\cos(\gamma^+)\sum_{j}(\delta\xi_{j}- \delta\xi_{i}) + J^-\sum_{j}\delta\eta_{j}\cos\left(\frac{4\pi}{N}(j-i)+ \gamma^-\right),\\
				\delta\dot{\eta_{i}} &= J^-\cos(\gamma^+)\sum_{j}(\delta\xi_{j}- \delta\xi_{i}) + J^+\sum_{j}\delta\eta_{j}\cos\left(\frac{4\pi}{N}(j-i)+ \gamma^-\right).
			\end{split}
			\label{eq:PWneg_perturb}
		\end{align}
	\end{widetext}
	The dynamics of the perturbations can again be arranged using a  $2N\times2N$ block matrix, and the stability of the system analyzed by its eigenvalues.  A detailed derivation of the eigenvalues is shown in Appendix~\ref{sec:appendix}. We obtain
	\begin{widetext}
		\begin{align}
			\begin{split}
				\lambda^{nPW}_0 &= 0 ,\\
				\lambda^{nPW}_1 &= -J_+\cos(\gamma^+),\\
				\lambda^{nPW}_{2\pm} & = \frac{J_+}{2} \left(\frac{1}{2}e^{-\iu\gamma^-} - \cos(\gamma^+)\right)\pm\frac{1}{2}\left[{J_+}^2\left(\frac{1}{2}e^{-\iu\gamma^-} + \cos(\gamma^+)\right)^2 - 2{J_-}^2e^{-\iu\gamma^-}\cos(\gamma^+)\right]^{1/2},\\
				\lambda^{nPW}_{(N-2)\pm} &= \frac{J_+}{2} \left(\frac{1}{2}e^{\iu\gamma^-} - \cos(\gamma^+)\right)\pm\frac{1}{2}\left[{J_+}^2\left(\frac{1}{2}e^{\iu\gamma^-} + \cos(\gamma^+)\right)^2 - 2{J_-}^2e^{\iu\gamma^-}\cos(\gamma^+)\right]^{1/2},
			\end{split}
			\label{eq:eig_nPWS}
		\end{align}
	\end{widetext}
	where $\lambda_{0}^{nPW}$ and $\lambda_{1}^{nPW}$ have multiplicities of $2$ and $(N-6)$, respectively. The superscript $PW$ stands for Phase Wave. Considering a positive position-phase correlation leads to slightly different eigenvalues. These differences, however, generate just a $\pi/2$ rotation of the stability regions (as will be shown in the next section). In Appendix~\ref{sec:appendix}, we summarize the derivation of the eigenvalues for the positively correlated Phase Wave states.
	
	For $J = K$ the non-zero eigenvalues are
	\begin{align}
		\begin{split}
			\lambda_{1}^{nPW (J = K)} &= - J_+\cos(\gamma^+),\\
			\lambda_{2+}^{nPW (J = K)} &= \frac{J_+}{2}e^{-\iu\gamma^-},\\
			\lambda_{2,(N-2)-}^{nPW (J = K)} &= -J_+\cos(\gamma^+),\\
			\lambda_{(N-2)+}^{nPW (J = K)} &= \frac{J_+}{2}e^{\iu\gamma^-},\\
		\end{split}
		\label{eq:simplifyPW}
	\end{align}
	and the regions where $\operatorname{Re}\{\lambda^{nPW(J = K)}\}$ are negative, for $J_+ > 0$ are $\gamma^+\in[-\pi/2 , \pi/2]$ and $\gamma^-\in[\pi/2, 3\pi/2]$, and for $J_+ < 0$ are $\gamma^+\in[\pi/2, 3\pi/2]$ and $\gamma^-\in[-\pi/2 , \pi/2]$. The regions for $J \neq K$, and for the positively correlated Phase Wave states are shown in the next section. Moreover, the symmetry $1/K \rightarrow K$ still applies for $J=1$, and so does the scaling in Eq.~\eqref{eq:eig_scaling}. The stability regions for $K$ and $1/K$ are, therefore, also the same for the Phase Wave states.

	\subsection{Asynchronous states}
	These states, such as the one shown in Figure~\ref{fig:static_states}c, are characterized by a uniform and uncorrelated distribution of particles in position and phase. To study their stability we take the limit of infinitely many oscillators and assume a continuum of particles described by the density function $\rho(x, \theta, t)\mathrm{d}x\mathrm{d}\theta$, which gives the fraction of particles lying between $x + \mathrm{d}x$ and $\theta + \mathrm{d}\theta$ at time $t$~\cite{strogatz1991stability}. The normalization condition for the density is 
	\begin{equation}
		\int_{0}^{2\pi}\int_{0}^{2\pi}\rho(x, \theta, t)\mathrm{d}x\mathrm{d}\theta = 1,
		\label{eq:norm_dens}
	\end{equation}
	which allows us to rewrite the order parameters, introduced in Eqs.~\eqref{eq:OP_fin}, as
	\begin{equation}
		S_\pm e^{\iu\phi_\pm} = \int_0^{2\pi}\int_0^{2\pi}e^{\iu(x \pm \theta)}\rho(x, \theta, t)\mathrm{d}x\mathrm{d}\theta.
		\label{eq:OP_inf}
	\end{equation}
	Given that the description of this state is more intuitive in terms of $x$ and $\theta$, we use Eqs.~\eqref{eq:frust_model} for its analysis. To reduce the size of the equations, however, we keep using the parameters $\xi$ and $\eta$ as defined before. Then, the equations of motion give the velocity vector field governing the behavior of the system:
	\begin{align}
		\begin{split}
			\dot{x} &= \frac{J}{2}S_+\sin(\phi_+ - \xi + \gamma^+) + \frac{J}{2}S_-\sin(\phi_- - \eta + \gamma^-) \\
			\dot{\theta} &= \frac{K}{2}S_+\sin(\phi_+ - \xi + \gamma^+) - \frac{K}{2}S_-\sin(\phi_- - \eta + \gamma^-).
		\end{split}
		\label{eq:model_cont}
	\end{align}
	
	The temporal evolution of the density $\rho(x, \theta, t)$ is described by the continuity equation
	\begin{equation}
		\frac{\partial \rho}{\partial t} + \nabla (\rho \vec{v}) = 0,
		\label{eq:conti}
	\end{equation}
	where $\vec{v} = (\dot{x} , \dot{\theta})$ as in Eqs.~\eqref{eq:model_cont}. The uniform density~$\rho_{0} = \pi^{-2}/4$ with $S_+=S_-=0$ is an equilibrium incoherent state and its stability can be studied by perturbing it as $\rho = \rho_0 + \delta\rho$ and analyzing the perturbation dynamics. The perturbation analysis is detailed in Appendix~\ref{sec:incoh}, and it leads to the derivation of the eigenvalues
	\begin{align}
		\begin{split}
			\lambda^{oAS}_{1\pm} &= \frac{J_+}{8\pi^2}e^{\pm\iu\gamma^+}, \\
			\lambda^{oAS}_{2\pm} &= \frac{J_+}{8\pi^2}e^{\pm\iu\gamma^-}.
			\label{eq:eigen_Async}
		\end{split}
	\end{align}
	The Static Asynchronous state will be stable when the real part of these eigenvalues are negative, that is, for $J_+>0$ when $\gamma^\pm\in[\pi/2 , 3\pi/2]$, and for $J_+<0$ when $\gamma^\pm\in[-\pi/2 , \pi/2]$.

	\section{Stability diagrams}
	\label{sec:stab_maps}
	
\begin{figure*}
		\centering
		\includegraphics[width = 0.9\textwidth]{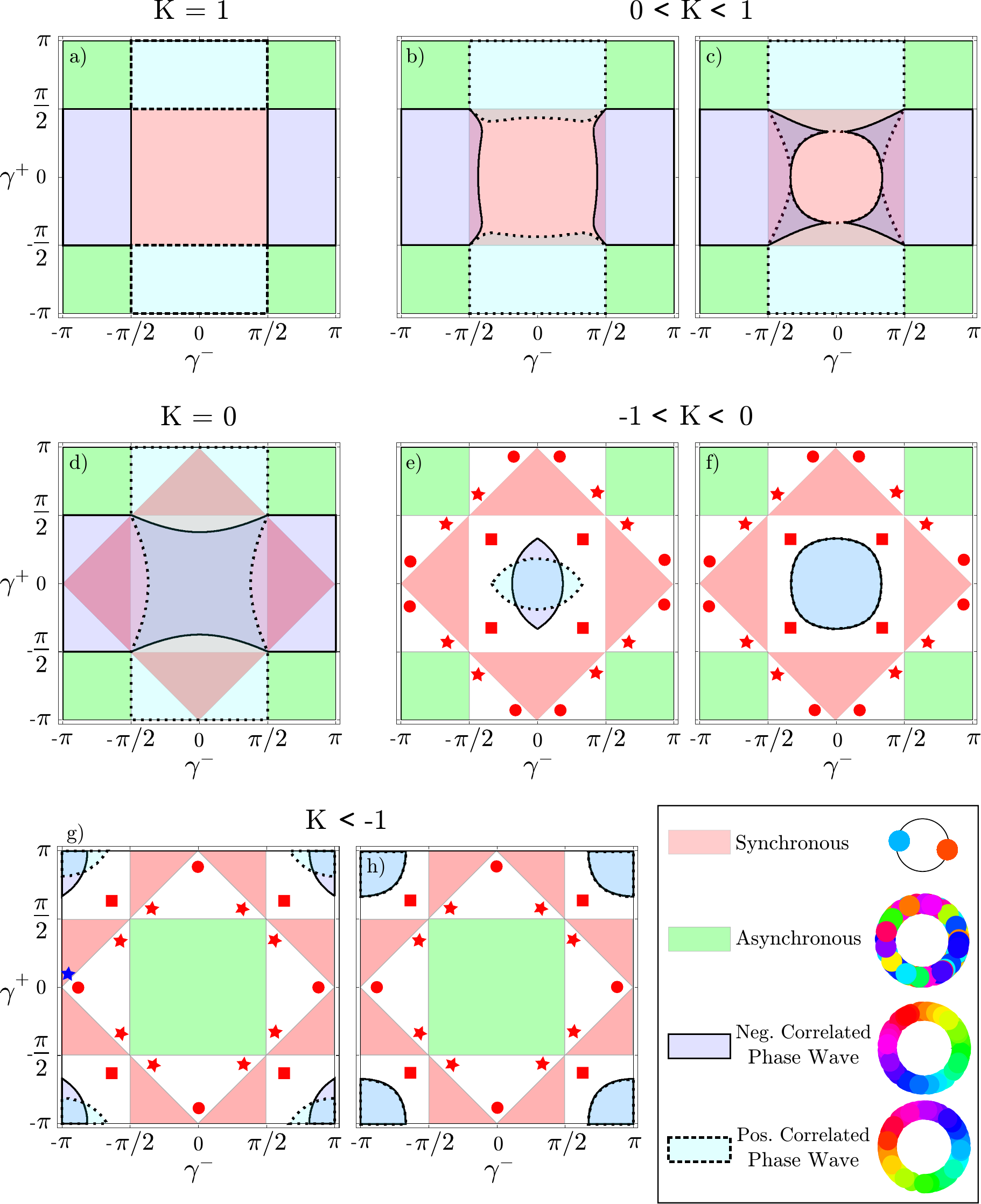}
		\caption{Stability regions computed using the eigenvalues obtained from the perturbation analyses. All figures were obtained for $J = 1$ and $N = 500$. In b), the diagram shows the regions for both $K = 1/5$ and $K = 5$. Similarly, in c), the diagram corresponds to both $K = 1/500$ and $K = 500$. In e) and f), $K = -1/5$ and $K = -1/500$, respectively. And, in g) and h), $K = -5$ and $K = -500$, respectively. Red circles, stars and squares are positioned in regions where states Turbulent, Noisy Active Phase Wave, and a combination of these two emerge, respectively. The blue star in g) corresponds to one instance of Ring state.} 
		\label{fig:stab_maps}
	\end{figure*}

	The  analytical results obtained in the previous section can be summarized with graphic representations of the Sakaguchi Swarmalators' stability regions spanned for $\gamma^\pm \in [-\pi$, $\pi]$. We fix $J = 1$ in all diagrams, as this specific setup allows us to observe the stability symmetry expected when using $K$ and $1/K$ (Figures~\ref{fig:stab_maps}b,~\ref{fig:stab_maps}c,~\ref{fig:stab_maps}e, and~\ref{fig:stab_maps}f). 
	
	The simplest case, where $J = K$, is shown in Figure~\ref{fig:stab_maps}a. As calculated in Eqs.~\eqref{eq:simplifySS}, ~\eqref{eq:simplifyPW}, and~\eqref{eq:eigen_Async}, the regions where each state emerges are just complementing squares. However, for $K > 0$ and different from $J$ (Figures~\ref{fig:stab_maps}b and~\ref{fig:stab_maps}c), the Phase Wave squared regions deform giving rise to the formation of stability regions that intersect. These are regions of multi-stability, where the system converges to either the Synchronous or Phase Wave states, depending on the initial conditions. Once $K$ becomes negative, the Synchronous square regions, depicted for $J \geq K$, split into four triangles, Figs.~\ref{fig:stab_maps}e and~\ref{fig:stab_maps}f, that split again for $K< -J$,  Figs.~\ref{fig:stab_maps}g, and~\ref{fig:stab_maps}h. Even more interesting is the behavior of the Phase Wave regions, which for $K$ approaching $-J$ from the right form intersecting $\pi/2$-rotated stripes (Figure~\ref{fig:stab_maps}e) that, when $K$ increases towards $0$, become fully intersected circles (Figure~\ref{fig:stab_maps}f). Similarly, the split of the Synchronous square and the Phase Wave stripes, which become circles, are found when $K$ is set below $-J$ (Figures~\ref{fig:stab_maps}g and~\ref{fig:stab_maps}h). Despite the similarities, however, these stability regions are $\pi$-translated from the $-J < K < 0$ regions described before, in both the $\gamma^-$ and $\gamma^+$ axes. We note that no such multi-stable regions exist in the Kuramoto-Sakaguchi model, suggesting that they depend on the interplay between the two degrees of freedom $\theta$ and $x$.
	
	A special case takes place when $K = -J$, which corresponds to $J_+ = 0$, and leads to the vanishing of negative eigenvalues in all states. Under these circumstances, the system exhibits the Active Asynchronous state, shown in Figure~\ref{fig:d_active_states}b. It is worth noting that, for these parameters, Eqs.~\eqref{eq:ring} reduce to a Hamiltonian system with $H=(J/2N)\sum_{i,j}\sin(x_j-x_i)\sin(\theta_j - \theta_i)$. However, for frustration parameters $(\alpha, \beta) \neq 0$, the system loses the Hamiltonian structure and, therefore, a constant of motion. The stability regions when $K = 0$ are shown in Figure~\ref{fig:stab_maps}d, and, despite these appearing to be well-defined, their nature is very susceptible to small numerical changes. The Noisy Phase Wave and Turbulent states (introduced in Figures~\ref{fig:d_active_states}a and~\ref{fig:d_active_states}c) are also found in the stability diagrams, as pointed by red markers in Figures~\ref{fig:stab_maps}e,~\ref{fig:stab_maps}f,~\ref{fig:stab_maps}g, and~\ref{fig:stab_maps}h. These, given their disordered nature, do not belong to any of the stable state's regions but are positioned in the blank spaces. Finally, the intriguing ring state, introduced in Figure~\ref{fig:o_active_states}c, turns out to be an intermediate state when entering the Synchronous states' stable regions (as depicted by the blue star in Figure~\ref{fig:stab_maps}g), which gives meaning to its clustered behavior. This state allows us to remark that, at the boundaries, linear stability analysis is not enough to decide the nature of the equilibrium.
	
\begin{figure*}
		\centering
		\includegraphics[width = 0.9\textwidth]{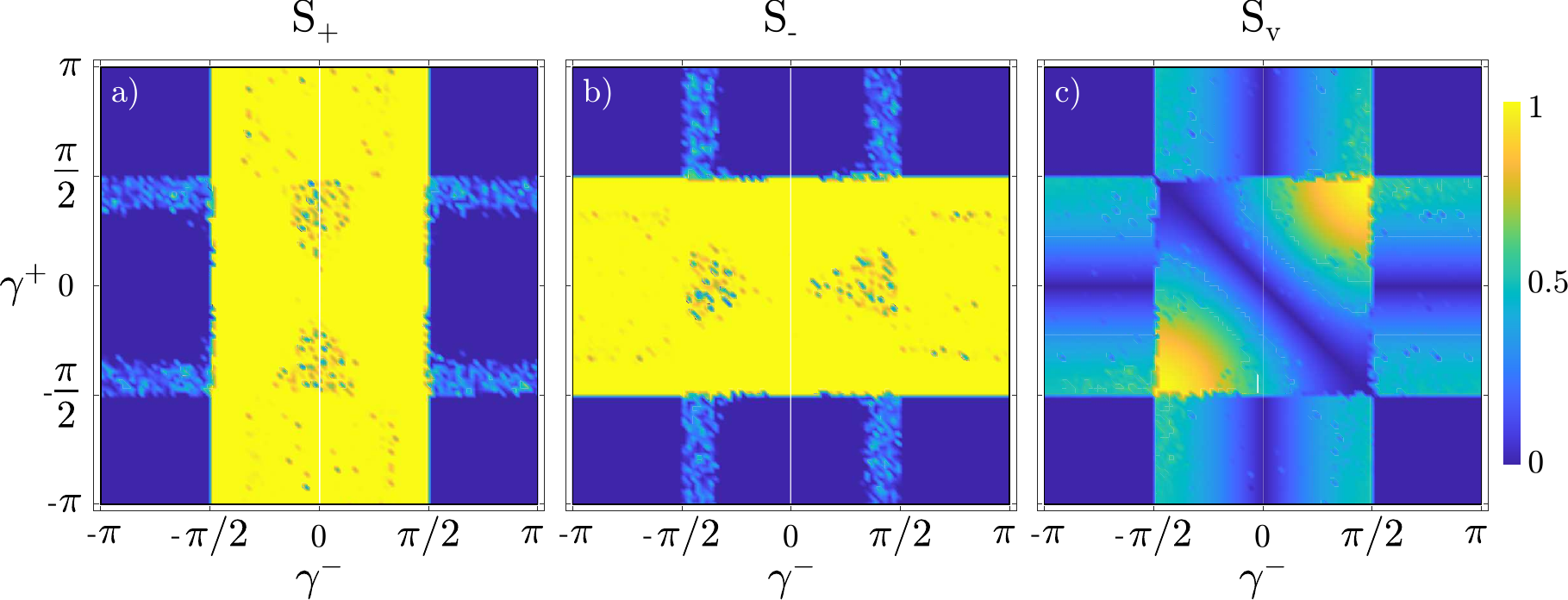}
		\caption{Order parameter heatmaps for $J = K = 1$ and $N = 500$, after $10^4$ time steps. Each intensity corresponds to the value of the respective order parameter for specific $(\gamma^+, \gamma^-)$ values. The diagrams show a) positive and b) negative position-phase correlation regions, and c) the average velocities.}
		\label{fig:op_profiles}
	\end{figure*}

	In Figure~\ref{fig:op_profiles}, we show heatmaps that complement the stability diagrams, presented in Figure~\ref{fig:stab_maps}, for the case where $J = K$. Each diagram is obtained for a system of $N = 500$ particles at its state after $10^4$ time steps. Intensities are positioned according to specific values of $\gamma^\pm$ used for the computation. In Figures~\ref{fig:op_profiles}a and~\ref{fig:op_profiles}b, these correspond to $S_+$ and $S_-$, respectively, calculated using Eqs.~\eqref{eq:OP_fin}. For Figure~\ref{fig:op_profiles}c, we use the additional parameter $S_v = 1/N\left|\sum_{j} {\dot{x}_j}\right|$, which is an indicator of the average velocity in the system. 
	
	Two main observations can be made by contrasting Figures~\ref{fig:stab_maps}a and~\ref{fig:op_profiles}. Firstly, as shown in the respective $S_\pm$ heatmaps, we can identify regions where the system converges to positively or negatively correlated states, which may represent either the Synchronous or Phase Wave states. Secondly, although convergence to any stable state can be proven by the eigenvalues, the definitions of $\xi$ and $\eta$ used for the stability analyses do not guarantee whether the system is static or active, even when analyzing $S_\pm$. However, the use of $S_v$ provides additional information that allows us to overcome these limitations. As shown in Figure~\ref{fig:op_profiles}c, clear partitions consistent with the regions in Figure~\ref{fig:stab_maps}a are displayed, and intensities reveal the velocity dependence on $\gamma^\pm$, which is coherent with the equilibrium trajectories defined in Section~\ref{sec:stab_analysis} for the Synchronous and Phase Wave states.
	
	\section{Conclusions}\label{sec:concl}
	We studied the effects of frustration induced disorder on a 1D Swarmalators system. Motivated by the work of Sakaguchi~\cite{sakaguchi1986soluble}, we modified the original system introduced in~\cite{o2022collective}, by including frustration parameters intended to break the coherence of the system in both position and phase spaces. 
	
	The most striking feature of the model is the emergence of active states for non-zero frustration parameters. These states remind us of the ones found in the 2D Swarmalators model~\cite{o2017oscillators}, that still lack a complete analytical explanation. In our model, however, we were able to find analytical solutions for the stability regions of each ordered state, independent of its static or active nature. Additionally, numerical computations allowed us to find regions where disordered active states emerge. In these states, despite the incoherent behavior exhibited by the particles, clear position-phase patterns can still be observed, which suggests that their analytical study could also be performed using different tools.
	
	From the stability analyses we see that, in contrast to the original 1D Swarmalators model, the frustration parameters provide us flexibility to find ordered states for any fixed $(J, K)$ values. That is, for a specific $(J, K)$ setup, we can find  Synchronous, positively or negatively correlated Phase Wave, or Asynchronous states just by tuning the values of $\gamma^\pm$. The disordered states, however, have been spotted only for  $K<0$. Even more exotic is the emergence of the Active Asynchronous case, which shows up only when $J = -K$.
	
	Although Eqs.~\eqref{eq:frust_model} have been defined in terms of an internal phase $\theta$ and a spatial coordinate $x$, we can think of the phase variable as another periodic spatial coordinate $y$, so that the scatter plots in Figures~\ref{fig:static_states},~\ref{fig:o_active_states}, and~\ref{fig:d_active_states} could represent particles' positions in the periodic Cartesian plane (a torus). The Active Asynchronous and Turbulent states are then similar to the patterns displayed by chiral rollers in~\cite{zhang2020reconfigurable} or ram semen in~\cite{schoeller2020collective}. The turbulent state, in particular, is of specific interest for future studies, as it shows the emergence of vortices and eddy-like structures (see movies S6 and S7 in Supplemental Material) even for finite number of particles.  Moreover, under this Cartesian setup, our results can also be extended to swarming-only systems whose position degrees of freedom interact.
	
	An interesting take on the Sakaguchi Swarmalators model would be to consider asymmetrical frustrations (i.e. parameters affecting only the sines or cosines), or even considering distributed couplings and frustrations, as done in~\cite{daido1992quasientrainment} for the Kuramoto model. In general, since the Kuramoto-Sakaguchi model and the concept of frustration have been widely studied~\cite{omel2012nonuniversal, manoranjani2021sakaguchi, arnaudon2022connecting, generalized}, the study of Sakaguchi Swarmalators can be expanded following these ideas.

	\begin{acknowledgments}
		It is a pleasure to thank Kevin P. O'Keeffe for helpful comments and suggestions. This work was partly supported by FAPESP grant 2021/14335-0 and CNPq grant 301082/2019‐7 (MAMA) and FAPESP grant 2021/04251-4 (JUFL). 
	\end{acknowledgments}
	
	\appendix
	
	\section{Eigenvalues for coherent states}\label{sec:appendix}
	The stability calculations, described in subsections~\ref{sub:SS} and~\ref{sub:PW}, show that perturbation dynamics can be arranged as 
	\begin{equation}
		\dot{\Vec{\delta_*}} = \mathbf{R}\Vec{\delta_*},
		\label{eq:arranged}
	\end{equation}
	where, for each system's state, the vector $\Vec{\delta_*}$ is composed of the individual perturbations $\delta\xi_i$ and $\delta\eta_i$, and
	\[\mathbf{R} = \begin{bmatrix}
		\mathbf{R_{11}} & \mathbf{R_{12}} \\
		\mathbf{R_{21}} & \mathbf{R_{22}}
	\end{bmatrix},\]
	is a matrix of circulant blocks. The structure of $\mathbf{R}$ allows us to find its eigenvalues $\lambda$ for all states, following the general procedure described below.
	
	The eigenvalues are solutions of the equation  
	\[\det(\mathbf{R} -\lambda\mathbf{I_{2N}}) = 0,\] 
	where $\mathbf{I_{2N}}$ is the identity matrix of dimension $2N$. However, since $[\mathbf{R_{11}}, \mathbf{R_{21}}] = 0$ holds for both the Synchronous and Phase Wave cases, we rewrite the equation for the determinant as
	\[\det(\mathbf{R} - \lambda\mathbf{I_{2N}}) = \det(\mathbf{M}),\]
	where $\mathbf{M} = (\mathbf{R_{11} - \lambda\mathbf{I_{N}}})(\mathbf{R_{22}- \lambda\mathbf{I_{N}}}) -  \mathbf{R_{21}}\mathbf{R_{12}}$ is also circulant. Then, the determinant of $\mathbf{M}$ can be computed using the general solution for circulant matrices
	\begin{equation}
		\det(\mathbf{M}) = \prod_{k = 0}^{N - 1}\left(M_{11} + M_{12}\zeta^{k} + \cdots + M_{1N}\zeta^{(N-1)k}\right),
		\label{eq:detM}
	\end{equation}
	where $\zeta$ is a primitive $N$-th root of unity, and the eigenvalues $\lambda$ are found by equating the resulting equations inside the parenthesis to zero.
	
	In the next subsections, we describe the solution to the eigenvalue problems for the Synchronous and Phase Wave states using the procedure described above. In each of these states, the blocks composing $\mathbf{R}$ have different structures that, however, can be considered special instances of circulant matrices. To differentiate between Synchronous and Phase Wave states, we use respectively the superscripts $^{SS}$ and $^{PW}$ on matrices and their elements. Since the Phase Wave state has two instances, we add the letters $n$ and $p$ in front of the superscripts to differentiate according to the negative or positive nature of the position-phase correlation. Additionally, to reduce the size of some expressions, we use $s_*$ and $c_*$ to represent $\sin(*)$ and $\cos(*)$ functions, respectively.
	
	\subsection{Synchronous states}
	Arranging Eqs.~\eqref{eq:ss_perturb} as Eq.~\eqref{eq:arranged} leads to a matrix $\mathbf{R}^{SS}$ composed by blocks 
	\begin{align*}
		\mathbf{R}^{SS}_\mathbf{{11}} &= J^+\cos(\gamma^+)\mathbf{R}_{\dagger},\\ \mathbf{R}^{SS}_\mathbf{{12}} &= J^-\cos(\gamma^-)\mathbf{R}_{\dagger},\\ \mathbf{R}^{SS}_\mathbf{{21}} &= J^-\cos(\gamma^+)\mathbf{R}_{\dagger},\\  \mathbf{R}^{SS}_\mathbf{{22}} &= J^+\cos(\gamma^-)\mathbf{R}_{\dagger},    
	\end{align*}
	where
	\begin{equation*}
		\mathbf{R}_{\dagger} = 
		\begin{pmatrix}
			1-N & 1 & \cdots & 1 \\
			1 & 1-N & \cdots & 1\\
			\vdots  & \vdots  & \ddots & \vdots  \\
			1 & 1 & \cdots & 1-N 
		\end{pmatrix}.
	\end{equation*}
	
	The composition of the block matrix $\mathbf{R}^{SS}$ allows us to infer that the off-diagonal terms of $\mathbf{M}^{SS}$ will all be the same. Thus, the only relevant elements to calculate $\det(\mathbf{M})$ are
	\begin{align*}
		M_{11}^{SS} &= \lambda^2 - \lambda J^+(1-N)(c_{\gamma^-}+ c_{\gamma^+}) \, + \\ 
		& \quad N(1-N)c_{\gamma^+}c_{\gamma^-}({J^-}^2 - {J^+}^2),\\
		M_{12}^{SS} &= -\lambda J^+ (c_{\gamma^-} + c_{\gamma+}) + Nc_{\gamma^+}c_{\gamma^-}({J^-}^2 - {J^+}^2),
	\end{align*}
	and Eq.~\eqref{eq:detM} can be rewritten as
	\begin{equation}
		\det(\mathbf{M}^{SS}) = \prod_{k = 0}^{N-1}\left(M_{11}^{SS} + M_{12}^{SS}\sum_{r = 1}^{N - 1}\zeta^{rk}\right),
		\label{eq:detSS}
	\end{equation}
	where \[\sum_{r = 1}^{N - 1}\zeta^{rk} = -1\]
	for $k\neq 0$. 
	
	Solving Eq.~\eqref{eq:detSS} leads to the product of $N$ quadratic equations, where $N-1$ of them are replicas. Equating this product to zero allows us to get the eigenvalue expressions shown in Eqs.~\eqref{eq:eigen_SS}.\\
	
	\subsection{Negatively correlated Phase Wave states}
	For these states, obtain the blocks
	\begin{align*}
		\mathbf{R}^{nPW}_\mathbf{{11}} &= J^+\cos(\gamma^+)\mathbf{R}_{\dagger},\\ \mathbf{R}^{nPW}_\mathbf{{12}} &= J^-\mathbf{R}^-_{*},\\ \mathbf{R}^{nPW}_\mathbf{{21}} &= J^-\cos(\gamma^+)\mathbf{R}_{\dagger},\\  \mathbf{R}^{nPW}_\mathbf{{22}} &= J^+\mathbf{R}^-_{*},    
	\end{align*}
	where
	\begin{widetext}
		\begin{equation*}
			\mathbf{R}^{\pm}_{*} = 
			\begin{pmatrix}
				\cos(\gamma^\pm) & \cos\left(4\pi\frac{1}{N}+ \gamma^\pm\right) & \cdots & \cos\left(4\pi\frac{N-1}{N}+ \gamma^\pm\right) \\
				\cos\left(4\pi\frac{N-1}{N}+ \gamma^\pm\right) & \cos(\gamma^\pm) & \cdots & \cos\left(4\pi\frac{N-2}{N}+ \gamma^\pm\right)\\
				\vdots  & \vdots  & \ddots & \vdots  \\
				\cos\left(4\pi\frac{1}{N}+ \gamma^\pm\right) & \cos\left(4\pi\frac{2}{N}+ \gamma^\pm\right) & \cdots & cos(\gamma^\pm) 
			\end{pmatrix}.
		\end{equation*}
	\end{widetext}
	
	Although we use only $\mathbf{R}^{+}_{*}$ to describe the $\mathbf{R}^{nPW}$ blocks, the matrix $\mathbf{R}^{-}_{*}$ will be used in the next subsection when describing the $\mathbf{R}^{pPW}$ blocks corresponding to the positive correlated Phase Wave states.
	
	The blocks in  $\mathbf{R}^{nPW}$ have different off-diagonal elements, which make the structure of $\mathbf{M}^{nPW}$ less intuitive. In this case the elementz needed to compute the determinant are
	\begin{widetext}
		\begin{align*}
			\begin{split}
				M_{11}^{nPW} &= \lambda^2 - \lambda J^+\left[c_{\gamma^-}+ (1-N)c_{\gamma^+}\right] + N\left({J^-}^2 - {J^+}^2\right)c_{\gamma^+}c_{\gamma^-},\\
				M_{1(r + 1)}^{nPW} &= -\lambda J^+\left[\cos\left(\frac{4\pi}{N}r + \gamma^-\right) + c_{\gamma^+}\right] + N\left({J^-}^2 - {J^+}^2\right)c_{\gamma^+}\cos\left(\frac{4\pi}{N}r + \gamma^-\right),
				\label{eq:MsPWS}
			\end{split}
		\end{align*}
	\end{widetext}
	%
	and Eq.~\eqref{eq:detM} can be rewritten as
	\begin{equation}
		\det(\mathbf{M}^{nPWS}) = \prod_{k = 0}^{N-1}\left(M_{11}^{nPW} + \sum_{r = 1}^{N-1}M_{1(r+1)}^{nPW}\zeta^{rk}\right).
		\label{eq:detPWS}
	\end{equation}
	
	Setting Eq.~\eqref{eq:detPWS} to zero should return the eigenvalues shown in~ Eqs.\eqref{eq:eig_nPWS}. However, since elements $M_{1(r + 1)}^{nPW}$ are all different, this solution is a bit more intricate. In Appendix~\ref{sec:simply}, we show the simplification of Eq.~\eqref{eq:detPWS} that allows us to get analytical solutions for the eigenvalues.
	
	\subsection{Positively correlated Phase Wave states}
	We start by summarizing the perturbation analysis since this was removed from the main text for the sake of clarity. For the Phase Wave states that exhibit a positive position-phase correlation, the equilibrium trajectories are
	\begin{align*}
		\xi_{i} &= J_{-}\sin({\gamma^-})t + \frac{4\pi i}{N} + \xi_{0},\\
		\eta_{i} &= J_{+}\sin({\gamma^-})t + \eta_{0}. 
	\end{align*}
	Adding perturbations to the equilibrium solutions we obtain
	\begin{align*}
		\begin{split}
			\delta\dot{\xi_{i}} &= J^-c_{\gamma^-}\sum_{j}(\delta\eta_{j}- \delta\eta_{i}) + J^+\sum_{j}\delta\xi_{j} C_{ij+},\\
			\delta\dot{\eta_{i}} &= J^+c_{\gamma^-}\sum_{j}(\delta\eta_{j}- \delta\eta_{i}) + J^-\sum_{j}\delta\xi_{j}C_{ij+},
		\end{split}
		\label{eq:PWpos_perturb}
	\end{align*}
	where $C_{ij+} = \cos\left(\frac{4\pi}{N}(j-i)+ \gamma^+\right)$. This
	can be arranged in blocks
	\begin{align*}
		\mathbf{R}^{pPW}_\mathbf{{11}} &= J^+\mathbf{R}^+_{*},\\ \mathbf{R}^{pPW}_\mathbf{{12}} &= J^-\cos(\gamma^-)\mathbf{R}_{\dagger},\\ \mathbf{R}^{pPW}_\mathbf{{21}} &= J^-\mathbf{R}^+_{*},\\  \mathbf{R}^{pPW}_\mathbf{{22}} &= J^+\cos(\gamma^-)\mathbf{R}_{\dagger}.    
	\end{align*}
	
	Despite the differences between $\mathbf{R}^{nPW}$ and $\mathbf{R}^{pPW}$, elements of $\mathbf{M}^{nPW}$ and $\mathbf{M}^{pPW}$ differ only by a swap of $\gamma^+$ and $\gamma^-$. Thus, we find
	\begin{widetext}
		\begin{align*}
			\begin{split}
				M_{11}^{pPW} &= \lambda^2 - \lambda J^+\left[(1-N)c_{\gamma^-} + c_{\gamma^+}\right] + N\left({J^-}^2 - {J^+}^2\right)c_{\gamma^+}c_{\gamma^-},\\
				M_{1(r + 1)}^{pPW} &= -\lambda J^+\left[c_{\gamma^-} + \cos\left(\frac{4\pi}{N}r + \gamma^+\right) \right] + N\left({J^-}^2 - {J^+}^2\right)\cos\left(\frac{4\pi}{N}r + \gamma^+\right)c_{\gamma^-}.
			\end{split}
		\end{align*}
	\end{widetext}
	Then, $\det(\mathbf{M}^{pPW})$ has the same form of Eq.~\eqref{eq:detPWS} and it can be solved using the simplification shown in Appendix~\ref{sec:simply}. The solution  leads to the eigenvalues
	\begin{widetext}
		\begin{align}
			\begin{split}
				\lambda^{pPW}_0 &= 0 ,\\
				\lambda^{pPW}_1 &= -J_+\cos(\gamma^-),\\
				\lambda^{pPW}_{2\pm} & = \frac{J_+}{2} \left(\frac{1}{2}e^{-\iu\gamma^+} - \cos(\gamma^-)\right)\pm\frac{1}{2}\left[{J_+}^2\left(\frac{1}{2}e^{-\iu\gamma^+} + \cos(\gamma^-)\right)^2 - 2{J_-}^2e^{-\iu\gamma^+}\cos(\gamma^-)\right]^{1/2},\\
				\lambda^{pPW}_{(N-2)\pm} &= \frac{J_+}{2} \left(\frac{1}{2}e^{\iu\gamma^+} - \cos(\gamma^-)\right)\pm\frac{1}{2}\left[{J_+}^2\left(\frac{1}{2}e^{\iu\gamma^+} + \cos(\gamma^-)\right)^2 - 2{J_-}^2e^{\iu\gamma^+}\cos(\gamma^-)\right]^{1/2}.
			\end{split}
		\end{align}
	\end{widetext}
	\section{Simplification}
	\label{sec:simply}
	We are interested in finding the solutions of a quadratic equation of the form 
	\[\lambda^2 + \lambda\phi + \omega = 0,\]
	where, from Eq.~\eqref{eq:detPWS}, 
	\begin{widetext}
		\begin{align*}
			\begin{split}
				\phi &= -J^+\left[c_{\gamma^-} +  c_{\gamma^+}\left( 1-N + \sum_{r = 1}^{N-1}\zeta^{rk}\right) +\sum_{r = 1}^{N-1} \zeta^{rk}\cos\left(\frac{4\pi}{N}r + \gamma^-\right) \right],\\
				\omega &= N({J^-}^2 - {J^+}^2)c_{\gamma^+}\left[c_{\gamma^-} + \sum_{r = 1}^{N-1} \zeta^{rk}\cos\left(\frac{4\pi}{N}r + \gamma^-\right)\right].
			\end{split}
		\end{align*}
	\end{widetext}
	We expand the sum in the right hand side of these equations as
	\begin{align*}
		\sum_{r = 1}^{N-1} \zeta^{rk} &\cos\left(\frac{4\pi}{N}r + \gamma^-\right) 
		=\frac{1}{2} e^{\iu\gamma^-}\sum_{r=1}^{N-1} e^{\iu r(4\pi + 2\pi k)/N}  \\ & + \frac{1}{2}e^{-\iu\gamma^-}\sum_{r = 1}^{N-1} e^{-\iu r(4\pi - 2\pi k)/N}.\\
	\end{align*}
	Then, for $ k \neq 2$,
	\begin{align*}
		\begin{split}
			\phi &= \begin{cases}
				0, & \text{if}\ k=0 \\
				J_+c_{\gamma^+}, & \text{otherwise}
			\end{cases},\\
			\omega &= 0,
		\end{split}
	\end{align*}
	which lead to pairs of eigenvalues where one of them is real and the other one is zero, or both are zero ($k = 0$). For $k = 2$,
	\begin{align*}
		\begin{split}
			\phi &= -J_+\left(\frac{1}{2}e^{-\iu\gamma^-} - c_{\gamma^+}\right),\\
			\omega &= \frac{1}{2}e^{-\iu\gamma^-}\left({J_-}^2 - {J_+}^2\right)c_{\gamma^+},
		\end{split}
	\end{align*}
	and $k = (N-2)$,
	\begin{align*}
		\begin{split}
			\phi &=  -J_+\left(\frac{1}{2}e^{\iu\gamma^-} - c_{\gamma^+}\right),\\
			\omega &= \frac{1}{2}e^{\iu\gamma^-}\left({J_-}^2 - {J_+}^2\right)c_{\gamma^+},
		\end{split}
	\end{align*}
	which lead to pairs of complex conjugate eigenvalues.
	
	\section{Perturbation analysis for incoherent states}
	\label{sec:incoh}
	Incoherent states are characterized by $S_\pm =0$ and, therefore, $\vec{v} = (\dot{x}, \dot{\theta}) = \vec{0}$. Thus, the homogeneous density $\rho_{0} = \pi^{-2}/4$ is a static solution of the continuity equation~\eqref{eq:OP_inf}.  
	
	Adding a small perturbation to the equilibrium state, $\rho = \rho_{0} + \delta\rho$, and using Eq.~\eqref{eq:conti}, we find that the temporal evolution of the perturbed state is governed by
	\begin{equation}
		\frac{\partial}{\partial{t}} {\delta\rho}= -\nabla \left( \delta\rho\right)\vec{v}, 
		\label{eq:pert_dyn_Async}
	\end{equation}
	where, from Eq.~\eqref{eq:norm_dens}
	\begin{equation}
		\int_{0}^{2\pi}\int_{0}^{2\pi}\delta\rho(x, \theta, t)\mathrm{d}x \mathrm{d}\theta = 0.
		\label{eq:pert_norm}
	\end{equation}
	To first order in $\delta\rho(x, \theta, t)$, we find that
	\begin{equation}
		S_{\pm}^1e^{\iu\phi_\pm} = \int_{0}^{2\pi}\int_{0}^{2\pi} e^{\iu (x\pm\theta)}\delta\rho(x, \theta, t)\mathrm{d}x\mathrm{d}\theta,
	\end{equation}
	which leads to 
	%
	\begin{align}
		\begin{split}
			\frac{\partial}{\partial t} \delta\rho &= \frac{J_+}{4\pi^2}\left(S^1_+ \cos(\phi_+ -\xi +\gamma^+) \right. \\
			& \left. + \, S^1_-\cos(\phi_- -\eta +\gamma^-)\right).
		\end{split}
		\label{eq:pertdyn_async}
	\end{align}
	%
	Expanding $\delta \rho$ in Fourier series 
	\begin{equation}
		\delta \rho = \sum_{m,n} f_{m,n}(t)e^{\iu(mx + n\theta)},
		\label{eq:fourier}
	\end{equation}
	and comparing with Eq.~\eqref{eq:pertdyn_async} 
	we see that the only relevant terms are $f_{\pm 1, \pm 1}$. We obtain 
	\begin{align}
		\begin{split}
			\dot{f}_{1,1}(t) &= \frac{J_+}{8\pi^2} e^{-\iu\gamma^-}f_{1,1}(t),\\
			\dot{f}_{-1,1}(t) &= \frac{J_+}{8\pi^2} e^{\iu\gamma^-}f_{-1,1}(t),\\
			\dot{f}_{1,-1}(t) &= \frac{J_+}{8\pi^2} e^{-\iu\gamma^+}f_{1,-1}(t),\\
			\dot{f}_{-1,-1}(t) &= \frac{J_+}{8\pi^2} e^{\iu\gamma^+}f_{-1,-1}(t).
		\end{split}
		\label{eq:fourier_eig}
	\end{align}
	Finally, writting $f(t) = \bar{f}e^{t\lambda}$, we can solve Eqs.~\eqref{eq:fourier_eig} to get the eigenvalues shown in Eqs.~\eqref{eq:eigen_Async}.

	
%

\end{document}